\documentclass[11pt]{article}

\usepackage{amsmath,amssymb,amsthm,amscd,latexsym}
\usepackage{hyperref}
\usepackage{cite}
\usepackage{graphicx,epsfig}
\RequirePackage{color}

\allowdisplaybreaks[1]

\def\nk{n_{\rm b}}

\def\rfr#1{eq. (\ref{#1})}

\def\derp#1#2{\rp{\partial{#1}}{\partial{#2}}}
\def\dert#1#2{\frac{{{d}}{#1}}{{{d}}{#2}}}

\def\virg#1{``#1''}

\def\eqi{\begin{equation}}
\def\eqf{\end{equation}}
\def\eqia{\begin{eqnarray}}
\def\eqfa{\end{eqnarray}}
\def\Om{\mathit{\Omega}}
\def\rp#1#2{{#1\over#2}}
\def\lb#1{\label{#1}}

\def\bds#1{\boldsymbol{#1}}

\def\co{\cos\omega}
\def\so{\sin\omega}

\def\sI{\sin I}

\def\ee{e^2}

\def\ton#1{\left(#1\right)}
\def\qua#1{\left[#1\right]}
\def\grf#1{\left\{#1\right\}}
\def\ang#1{\left\langle #1\right\rangle}

\begin{document}

\title{Constraints on a  MOND effect for  isolated aspherical systems in deep Newtonian regime from orbital motions}

\author{L. Iorio\\ Ministero dell'Istruzione, dell'Universit$\grave{\textrm{a}}$ e della Ricerca (M.I.U.R.)-Istruzione \\ Fellow of the Royal Astronomical Society (F.R.A.S.)\\ Viale Unit$\grave{\textrm{a}}$ di Italia 68, 70125, Bari (BA), Italy}

\maketitle

\begin{abstract}
The dynamics of non-spherical systems described by MOND theories arising from generalizations of the Poisson equation is affected by an extra MONDian quadrupolar potential $\phi_{\rm Q}$ even if they are isolated (no EFE effect) and if they are in deep Newtonian regime. In general MOND theories quickly approaching Newtonian dynamics for accelerations beyond $A_0$, $\phi_{\rm Q}$ is proportional to a coefficient $\alpha\sim 1$, while in MOND models becoming Newtonian beyond $\kappa A_0,\ \kappa\gg 1,$  it is enhanced by $\kappa^2$. We analytically work out some orbital effects due to $\phi_{\rm Q}$ in the framework of QUMOND, and compare them with the latest observational determinations of Solar System's planetary dynamics, exoplanets, double lined spectroscopic binary stars and binary radio pulsars. The current admissible range for the anomalous perihelion precession of Saturn \textcolor{black}{$-0.5$ milliarcseconds per century $\leq \Delta\dot\varpi\leq 0.8$ milliarcseconds per century} yields $|\kappa|\leq \textcolor{black}{3}.5\times 10^{\textcolor{black}{3}}$, while the radial velocity of $\alpha$ Cen AB allows to infer $|\kappa|\leq 6.2\times 10^4\ ({\rm A})$ and  $|\kappa|\leq 4.2\times 10^4\ ({\rm B})$.
\end{abstract}



\centerline
{PACS: 04.80.-y; 04.80.Cc; 95.1O.Ce;  95.10.Km; 97.80.-d}

\section{Introduction}
The MOdified Newtonian Dynamics (MOND) (see \cite{lrr-2012-10} for a recent review) is a theoretical framework proposed by Milgrom \cite{1983ApJ...270..365M,1983ApJ...270..371M,1983ApJ...270..384M} to modify the laws of the gravitational interaction in a suitably defined low acceleration regime to explain  the observed anomalous kinematics of certain astrophysical  systems such as various kinds of galaxies \cite{1980ApJ...238..471R,1995AJ....109..151V,2008AJ....136.2563W}. Indeed, their behaviour does not agree with the predictions made with the usual Newtonian inverse-square law applied to the electromagnetically detected baryonic matter whose quantity appears to be insufficient. \textcolor{black}{In the case of the mass discrepancy occurring in clusters of galaxies \cite{1933AcHPh...6..110Z}, MOND actually faces difficulties in explaining it \cite{1999ApJ...512L..23S,2008MNRAS.389..250N,2009MNRAS.396..887A}.} In almost all its relativistic formulations, MOND implies a single\footnote{An exception is TeVeS \cite{2004PhRvD..70h3509B}, in its original form.} acceleration scale \cite{1991MNRAS.249..523B} $A_0=(1.2\pm 0.27)\times 10^{-10}$ m s$^{-2}$ \textcolor{black}{below} which the laws of gravitation would suffer notable modifications mimicking the effect of the additional non-baryonic Dark Matter which is usually invoked to explain the observed discrepancy  within the standard theoretical framework.

\textcolor{black}{
In this paper, we propose to constrain a recently predicted strong-field effect of MOND \cite{2012MNRAS.426..673M} by using various observables pertaining different astronomical scenarios. In the following we will briefly outline the main features of such a novel prediction of MOND which occurs even if the system under consideration is isolated and if its characteristic accelerations are quite larger than $A_0$.}

\textcolor{black}{Let us consider an isolated, strongly gravitating system $\mathcal{S}$ of total mass $M_{\rm tot}$ and extension \eqi R\ll d_{\rm M}\doteq\sqrt{\rp{GM_{\rm tot}}{A_0}},\eqf where $G$ is the Newtonian constant of gravitation. Let us also assume that the mass distribution of $\mathcal{S}$, characterized by a generally anisotropic matter density $\varrho(\bds r)$,  varies  over timescales much larger than $t_{\rm M}\doteq d_{\rm M}/c,$ where $c$ is the speed of light in vacuum.
According to formulations of MOND based on extensions of the Poisson equation such as the nonlinear Poisson model by Bekenstein and Milgrom \cite{1984ApJ...286....7B} and\footnote{\textcolor{black}{It is the nonrelativistic limit of a certain formulation of bimetric MOND (BIMOND) \cite{2009PhRvD..80l3536M}.}} QUMOND \cite{2010MNRAS.403..886M}, it turns out \cite{2012MNRAS.426..673M} that, if on the one hand,  the MOND field equations of $\mathcal{S}$ coincide with the usual linear Poisson equation for $r\leq R$ depending on how the MOND interpolating function $\mu$ is close to unity, on the other hand, they differ from it for $r\geq d_{\rm M}$. This is a crucial feature since it implies that the solution $\phi$ of the Poisson equation for $r\leq R$ is, in general, different from the usual Newtonian one $\phi_{\rm N}$, thus affecting the internal dynamics of $\mathcal{S}$ even if it is in the strong gravity regime. It is as if a hollow \virg{phantom} matter distribution, characterized by a phantom matter density $\varrho_{\rm ph}(\bds r)$, was present at $r\geq d_{\rm M}$ in such a way that, in the quasi-static limit previously defined,  $\varrho(\bds r)$ instantaneously controls $\varrho_{\rm ph}(\bds r)$ by fixing its symmetry properties. If $\varrho=\varrho(r)$, i.e. if $\mathcal{S}$ is spherically symmetric, then the phantom matter density is spherically symmetric as well. In this case, the internal dynamics of  $\mathcal{S}$ would not be affected by the peculiar boundary conditions on the MOND field equations at $r\geq d_{\rm M}$ or, equivalently, by the phantom matter. Indeed, it would be arranged in a hollow spherical shell; the dynamics of $\mathcal{S}$ would be Newtonian to the extent that the MOND interpolating function $\mu$ matches the unity. On the contrary, if   $\varrho=\varrho(\bds r)$, i.e. if $\mathcal{S}$ is not spherically symmetric, the same occurs to the phantom matter as well. Thus, it does have an influence on the internal dynamics of $\mathcal{S}$ which, to the lowest order, can be approximated by an additional quadrupolar potential\footnote{\textcolor{black}{The difference $\phi-\phi_{\rm N}$  can be thought as the solution of usual linear Poisson equation just for the phantom matter density $\varrho_{\rm ph}$ \cite{2012MNRAS.426..673M}.}} $\phi_Q = \phi-\phi_{\rm N}$.
}

By assuming $\mu=1$ to the desired accuracy everywhere within \textcolor{black}{$\mathcal{S}$} and by using QUMOND \cite{2010MNRAS.403..886M}, Milgrom \cite{2012MNRAS.426..673M} obtained
\eqi \phi_{Q}({\bds r}_F) = -\rp{\alpha G}{\textcolor{black}{d}^5_{\rm M}}x_F^{i} x_F^{j}Q_{ij}\lb{Upot}\eqf
with
\eqi Q_{ij}\doteq \rp{1}{2}\int_{\mathcal{S}} \varrho\ton{{\bds r}^{'}}\ton{r^{' 2}\delta_{ij}-3x_i^{'} x_j^{'}}d{\bds r}^{'}\lb{qpot};\eqf
 $x_F^i,i=1,2,3$ in \rfr{Upot} are the components of the position vector ${\bds r}_F$ of a generic point \textit{F}  with respect to \textcolor{black}{the barycenter} of \textcolor{black}{$\mathcal{S}$}, while $x^{'}_j, j=1,2,3$ in \rfr{qpot} determine the \textcolor{black}{barycentric} position of the system's mass elements. \textcolor{black}{The coefficient $\alpha$ depends on the specific form of the interpolating function chosen. Milgrom \cite{2012MNRAS.426..673M}, by considering also the case in which the strong field regime is obtained in terms of a second, dimensionless constant $\kappa\gg 1$ when the Newtonian acceleration is as large as $\sim \kappa A_0$, picked up an interpolating function yielding
\eqi\alpha_{\kappa}=\kappa^2 \alpha_{\kappa=1},\ \alpha_{\kappa=1}\sim 1.\lb{limite}\eqf   }
\textcolor{black}{In general, there should be many other interpolating functions that could be used with $\kappa\gg 1$; in this paper, we will focus on \rfr{limite}.}
Finally, we remark that Milgrom \cite{2012MNRAS.426..673M} felt that theories with $\kappa\gg 1$ cannot be considered as generic MOND results.

As stressed by Milgrom \cite{2012MNRAS.426..673M}, the quadrupolar MOND effect of \rfr{Upot}  has not to be confused with some other MONDian features occurring in the strong acceleration regime which were previously examined in literature. In particular, it is not the quadrupolar effect \cite{2009MNRAS.399..474M,2011MNRAS.412.2530B} due to the External Field Effect (EFE) \cite{1983ApJ...270..365M,1984ApJ...286....7B,1986ApJ...302..617M} arising when the system under consideration is immersed in an external background field; indeed, here the system is considered isolated. Even so,  residual MONDian effects in the strong acceleration regime exist, in general, because of the remaining departure of $\mu(q)$ from 1 when $q\gg 1$; their consequences on orbital motions of Solar System objects were treated in, e.g., \cite{1983ApJ...270..365M,2006MNRAS.371..626S,2008JGrPh...2...26I,2009MNRAS.399..474M}. Nonetheless, they are different from the presently studied effect, for which it was posed $\mu=1$ to the desired accuracy. Finally, Milgrom \cite{2012MNRAS.426..673M} showed that the impact  of the zero-gravity points \cite{2006PhRvD..73j3513B,2012PhRvD..86d4002G,2012PhRvD..85d3527M} existing in high acceleration regions  on the dynamics of the mass sources themselves is negligible with respect to the effect considered here.

The plan of the paper is as follows. In Section \ref{calcolo} we analytically work out some orbital effects caused by \rfr{Upot} to an isolated two-body system in the case of \rfr{limite}. In Section \ref{osservazione} our results are compared to latest observations on Solar System planetary motions, extrasolar planets, and spectroscopic binary stars. Section \ref{sommario} is devoted to summarizing our findings.
\section{Calculation of some orbital effects}\lb{calcolo}
\textcolor{black}{Let us consider} a typical non-spherical system such as a localized binary  made of two point masses $M$ and $m$ with $M_{\rm b}\doteq M_{\rm tot} = M+m$. \textcolor{black}{In a barycentric frame,} its mass density \textcolor{black}{$\varrho\ton{{{\bds r}_{F}}}$} at a generic point \textit{F}  can be posed
\eqi\textcolor{black}{\varrho}\ton{{{\bds r}_{F}}}=M\delta^3\ton{{{\bds r}_F} - {\bds r}_M} + m\delta^3\ton{{{\bds r}_F} - {\bds r}_m}\lb{rho},\eqf where ${\bds r}_m$ and ${\bds r}_M$ \textcolor{black}{are the barycentric} position vectors of $m$ and $M$, respectively.
%
%
%
%
%
%
%
%
%
%

\textcolor{black}{After having calculated $\phi_Q({\bds r}_F)$ for \rfr{rho}, its gradient with respect to ${\bds r}_F$ yields the extra-acceleration ${\bds A}_F$ of an unit  mass at a generic point $F$. The extra-accelerations ${\bds A}_m$ and ${\bds A}_M$ experienced by $m$ and $M$ can be obtained by calculating ${\bds A}_F$ for ${\bds r}_m= M M_{\rm b}^{-1}\bds r$ and for ${\bds r}_M= -m M_{\rm b}^{-1}\bds r$, respectively, where $\bds r\doteq {\bds r}_m - {\bds r}_M$ is the relative position vector directed from $M$ to $m$.}
%
%
%
%
\textcolor{black}{It turns out that the accelerations felt by $m$ and $M$ are
\begin{align}
{\bds A}_m \lb{Accelm} &= -\rp{2\alpha A_0 m M^2}{M^3_{\rm b}d^3_{\rm M}}r^2\bds r, \\ \nonumber \\
{\bds A}_M \lb{AccelM} &= \rp{2\alpha A_0 M m^2}{M^3_{\rm b}d^3_{\rm M}}r^2\bds r.
\end{align}
The relative extra-acceleration is, thus, \cite{2012MNRAS.426..673M}
\eqi \bds A = -\rp{2\alpha A_0 }{d^3_{\rm M}}\ton{\rp{\mu_{\rm b}}{M_{\rm b}}}r^2\bds r,\lb{AQ}\eqf
where $\mu_{\rm b}\doteq m M M_{\rm b}^{-1}$ is the binary's reduced mass.}
For the following developments, it is useful to remark that, formally, \rfr{AQ} can be derived from the effective potential
\eqi U_{\rm M}=\rp{\alpha A_0 }{2d^3_{\rm M}}\ton{\rp{\mu_{\rm b}}{M_{\rm b}}}r^4\lb{UQ}.\eqf
\subsection{The pericenter rate \textcolor{black}{for} \textcolor{black}{a} \textcolor{black}{two-body} \textcolor{black}{MOND} \textcolor{black}{quadrupole}}
The longitude of pericenter $\varpi\doteq \mathit{\Omega} + \omega$ is a \virg{broken} angle since the longitude of the ascending node $\Om$ lies in the reference $\grf{x,y}$ plane from the reference $x$ direction to the line of the nodes\footnote{It is the intersection of the orbital plane with the reference $\grf{x,y}$ plane.}, while the argument of pericenter $\omega$ reckons the position of the point of closest approach in the orbital plane with respect to the line of the nodes. The angle $\varpi$ is usually  adopted in Solar System  studies to put constraints on putative modifications of standard Newtonian/Einsteinian dynamics \cite{2011CeMDA.111..363F}.  Its Lagrange perturbation equation is \cite{befa}
\eqi\ang{\dert\varpi t} = -\rp{1}{n_{\rm b}a^2}\qua{\ton{\rp{\sqrt{1-\ee}}{e}}\derp{\ang{U_{\rm pert}}} e + \rp{\tan\ton{\rp{I}{2}}}{\sqrt{1-\ee}}\derp{\ang{U_{\rm pert}}} I },\lb{lagvarpi}\eqf where $U_{\rm pert}$ is a small correction to the Newtonian potential;  $a$ is the relative semimajor axis, $e$ is the orbital eccentricity, and $I$ is the inclination of the orbital plane to the reference $\grf{x,y}$ plane. The brackets $\ang{\ldots}$ in \rfr{lagvarpi} denote the average over one full orbital period $P_{\rm b}=2\pi \nk^{-1}=2\pi\sqrt{a^3 G^{-1} M_{\rm b}^{-1}}$. By adopting \rfr{UQ} as perturbing potential $U_{\rm pert}$ in \rfr{lagvarpi}, one gets
\eqi \ang{\dert\varpi t}  = -\rp{5\alpha A_0 P_{\rm b}}{2\pi d_{\rm M}}\ton{\rp{\mu_{\rm b}}{M_{\rm b}}}\ton{\rp{a}{d_{\rm M}}}^2\sqrt{1-\ee}\ton{1 + \rp{3}{4}\ee}\lb{rateQ}. \eqf
\textcolor{black}{It should be remarked that \rfr{UQ} and, thus, \rfr{rateQ} are valid just for a two-body MOND quadrupole $Q_{ij}$. }

As a cross-check of the validity of our result, we repeated the calculation of the long-term precession of $\varpi$ by using \rfr{AQ} as perturbing acceleration and  the Gauss equations for the variations of the elements: we re-obtained \rfr{rateQ}.
\subsection{The timing in binary radi\textcolor{black}{o}pulsars}
The basic observable in  binary pulsar systems is the \textcolor{black}{periodic} change $\delta\tau_{\rm p}$ in the time of arrivals (TOAs) $\tau_{\rm p}$ of the pulsar p \textcolor{black}{due to the fact that it is gravitationally bounded to a generally unseen companion c, thus describing an orbital motion around the common barycenter}. \textcolor{black}{
In a binary hosting an emitting radiopulsar, the Keplerian expression of $\delta\tau_{\rm p}$ is obtained by taking the ratio of the component $\rho_{\rm p}$ of the barycentric pulsar's orbit along the line of sight to the speed of light $c$. Thus, one has \eqi\delta\tau_{\rm p}=\rp{\rho_{\rm p}}{c}.\lb{fesso}\eqf Since the line of sight is customarily assumed as reference $z$ axis, in \rfr{fesso} it is  \eqi \rho_{\rm p}\equiv z_{\rm p},\ z_{\rm p}=r_{\rm p}\sI\sin(\omega+f),\lb{zetap}\eqf as it can be inferred from the standard expressions for the orientation of the Keplerian ellipse in space. In \rfr{zetap}, $r_{\rm p}$ is the distance of the pulsar from the system's center of mass, $I$ is the inclination of the orbit to the plane of the sky, assumed as reference $\{x,y\}$ plane, and $f$ is the true anomaly reckoning the instantaneous position of the pulsar with respect to the periastron position. By using
\begin{align}
r_{\rm p} &= a_{\rm p}(1 - e \cos E), \\ \nonumber \\
\cos f & = \rp{\cos E -e}{1 - e \cos E}, \\ \nonumber \\
\sin f & = \rp{\sqrt{1-e^2}\sin E}{1 - e \cos E},
\end{align}
where $a_{\rm p}$ is the semimajor axis of the the pulsar's barycentric orbit and $E$ is the eccentric anomaly, from \rfr{fesso}-\rfr{zetap} one straightforwardly gets
}
\cite{Dam91,2000ApJ...544..921K}
\eqi\delta\tau_{\rm p} = {\rm x}_{\rm p}\qua{\ton{\cos E - e}\so + \sqrt{1-e^2}\sin E\co}.\lb{tempop}\eqf
 \textcolor{black}{In \rfr{tempop},} ${\rm x}_{\rm p}\doteq a_{\rm p}\sI/c$  is the projected semimajor axis of the pulsar's barycentric orbit \textcolor{black}{and has dimensions of time}; \textcolor{black}{by posing} $m_{\rm p}\doteq M,m_{\rm c}\doteq m$, it is $a_{\rm p}\textcolor{black}{\doteq}a_M= m M^{-1}_{\rm b} a,$ \textcolor{black}{where $a$ is the semimajor axis of the pulsar-companion relative orbit}.

In general, the \textcolor{black}{shift per orbit} $\Delta Y$ of an observable $Y$  \textcolor{black}{with respect to its classical expression due to the action of a perturbing acceleration such as either \rfr{Accelm} or \rfr{AccelM} } can be computed as
\eqi \Delta Y=\int_0^{P_{\rm b}}\ton{\dert Y t}dt =\int_0^{2\pi}\qua{\derp Y{E}\dert {E}{\mathcal{M}}\dert{\mathcal{M}}t + \sum_{\psi}\derp Y\psi\dert\psi t}\ton{\dert t{E}}dE,\lb{integrale}\eqf where ${\mathcal{M}}$ is the mean anomaly and $\psi$ collectively denotes the other Keplerian orbital elements. The rates $\dot{\mathcal{M}},\dot\psi$  entering \rfr{integrale} \textcolor{black}{are due to the perturbation and} are instantaneous. \textcolor{black}{As such, they} are obtained by computing the right-hand-sides of \textcolor{black}{either} the Lagrange equations \textcolor{black}{or the Gauss equations} onto the unperturbed Keplerian ellipse without averaging them over $P_{\rm b}$. \textcolor{black}{The derivatives $\partial Y/\partial E,\partial Y/\partial\psi$ in \rfr{integrale} are computed by using the unperturbed expression for $Y$.}

By using \rfr{integrale} \textcolor{black}{and
\eqi \rp{dt}{dE}=\rp{1-e\cos E}{n_{\rm b}},\eqf }  the MOND time shift perturbation can be computed as
\eqi\Delta\delta\tau_{\rm p} = -\rp{7\alpha A_0 P^2_{\rm b}}{8\pi c}\ton{\rp{m\mu_{\rm b}}{M_{\rm b}^2}}\ton{\rp{a}{d_{\rm M}}}^3  e\sqrt{1-\ee} \ton{1+\rp{\ee}{2}}\co\sI.\lb{taup}\eqf
\textcolor{black}{It is important to notice that \rfr{taup} is proportional to $P^2_{\rm b}$ and to $e$. At a first sight, it may be weird to see in \rfr{taup} a dependence on the speed of light $c$ in a non-relativistic theory such as QUMOND; actually, it is not so because of the definition of $\tau_{\rm p}$ in \rfr{fesso}.  }
\subsection{The radial velocity}
The radial velocity $V_{\rho_{\rm lc}}$ \cite{Latham00} is a standard observable in spectroscopic studies of binaries \cite{Batten00}. Up to the radial velocity of the binary's center of mass $V_0$, the Keplerian expression of the radial velocity of the component of the binary  whose  light curve (lc) is available \textcolor{black}{can be obtained by taking the time derivative of the projection $\rho_{\rm lc}$ of the barycentric orbit of the visible component  onto the line of sight. Thus, from \rfr{zetap}, it can be posed \eqi V_{\rho_{\rm lc}}=\dert{\rho_{\rm lc}}{t}\equiv\dert{z_{\rm lc}}t=\derp{z_{\rm lc}}{f}\derp{f}{{\mathcal{M}}}\nk.\lb{piffo}\eqf
By using the standard Keplerian expressions
\begin{align}\derp{f}{{\mathcal{M}}} & = \ton{\rp{a_{\rm lc}}{r_{\rm lc}}}^2\sqrt{1-e^2},\\ \nonumber \\
r_{\rm lc} & = \rp{a_{\rm lc}(1-e^2)}{1+e\cos f},
\end{align} where $r_{\rm lc}$  and $a_{\rm lc}$
refer to the barycentric orbit of the visible partner,
\rfr{piffo} straightforwardly yields
 }
\eqi V_{\rho_{\rm lc}} = K\qua{e\co + \cos\ton{\omega+f}}=\rp{\nk a_{\rm lc}\sI}{\sqrt{1 - \ee}}\qua{e\co + \cos\ton{\omega+f}}.\lb{velr}\eqf \textcolor{black}{In \rfr{velr},} $K$ is the semi-amplitude of the radial velocity. In the case of extrasolar planetary systems, \textcolor{black}{the light curve  is usually available only for} the hosting star; thus, $a_{\rm lc}\textcolor{black}{\doteq}a_M = m M^{-1}_{\rm b}a$. In the case of spectroscopic binary stars, it may happen that the light curves of both the components (double lined spectroscopic binary stars) are available.

\textcolor{black}{As for $\Delta\delta\tau_{\rm p}$, also the perturbation $\Delta V_{\rho_{\rm lc}}$ of the radial velocity due to a disturbing extra-acceleration can be calculated from \rfr{integrale}. In this case, it is computationally more convenient to replace $E$ with $f$ throughout \rfr{integrale}; as a consequence,
\eqi \dert t f = \rp{\ton{1-e^2}^{3/2}}{\nk\ton{1+e\cos f}^2}\eqf
must be used. }
The MOND  perturbation of $V_{\rho_{\rm lc}}$ turns out to be
\eqi \Delta V_{\rho_{\rm lc}} = \rp{17}{2}\alpha A_0 P_{\rm b}\ton{\rp{m\mu_{\rm b}}{M_{\rm b}^2}}\ton{\rp{a}{d_{\rm M}}}^3 e\ton{1 + \rp{8}{17}\ee} \sI\so.\lb{dvr}\eqf
\textcolor{black}{It is important to note the proportionality of \rfr{dvr} to $P_{\rm b}$ and to $e$.}
\section{Confrontation with the observations}\lb{osservazione}
\subsection{Planets of the Solar System}\lb{pianeta}
As far as the Solar System is concerned, $t_{\rm M}=39$ d; thus the quasi-staticity condition is fully satisfied by the gaseous giant planets for which it is $P_{\rm b}\gtrsim 4300$ d.

Among them, Saturn, whose orbital period is as large as $P_{\rm b}=10759$ d, is the most suitable to effectively constrain $\alpha$ since its orbit is nowadays known with $\approx 20$ m accuracy \cite{2011CeMDA.111..363F} in view of the multi-year record of accurate radio-technical data from the Cassini spacecraft. Looking at its perihelion, any deviation of its secular precession from \textcolor{black}{the rate predicted by the standard Newtonian/Einsteinian dynamics} can  nowadays be constrained down to sub-milliarcseconds per century (mas cty$^{-1}$) level, as shown by Table \ref{tavolaf}.
\begin{table*}[ht!]
\caption{Supplementary precessions $\Delta\dot{\mathit{\Omega}}, \Delta\dot \varpi$ of
the longitudes of the node  and of the perihelion for some planets of the Solar System 
 estimated by Fienga et al. \cite{2011CeMDA.111..363F} with the INPOP10a ephemerides.
 Data from Messenger and Cassini were used. Fienga et al. \cite{2011CeMDA.111..363F} fully modeled all standard Newtonian/Einsteinian dynamics, apart from the Solar Lense-Thirring effect, which, however, is relevant only for Mercury\textcolor{black}{; MOND was not modelled}.
 The reference $\{x,y\}$ plane is the mean Earth's equator at J$2000.0$. The units are  milliarcseconds per century (mas cty$^{-1}$).
}\label{tavolaf}
\centering
\bigskip
\begin{tabular}{lll}
\hline\noalign{\smallskip}
&   $\Delta\dot \Om$ (mas cty$^{-1}$) & $\Delta\dot \varpi $ (mas cty$^{-1}$)  \\
\noalign{\smallskip}\hline\noalign{\smallskip}
Mercury & $1.4 \pm 1.8$ & $0.4 \pm 0.6$ \\
Venus & $0.2 \pm 1.5$ & $ 0.2\pm 1.5$ \\
Earth & $0.0\pm 0.9$ & $-0.2\pm 0.9$ \\
Mars & $-0.05\pm 0.13$ & $-0.04\pm 0.15$ \\
%
%
Saturn & $-0.1\pm 0.4$ & $0.15\pm 0.65$ \\
\noalign{\smallskip}\hline\noalign{\smallskip}
\end{tabular}
\end{table*}

If the case $\alpha_{\kappa}=\kappa^2 \alpha_{\kappa=1}$, with $\alpha_{\kappa=1}\sim 1$ is considered, \textcolor{black}{the two-body expression of} \rfr{rateQ} and Table \ref{tavolaf}
yield
\eqi|\kappa| \leq 2.5\times 10^{5} \lb{satc};\eqf \textcolor{black}{larger values for $|\kappa|$ would yield an anomalous secular perihelion precession exceeding the allowed bounds in Table \ref{tavolaf}.}

\textcolor{black}{Actually, our analysis is incomplete since it is limited to a two-body scenario. As remarked by Milgrom himself \cite{2012MNRAS.426..673M}, also the contribution of the other planets, especially the more massive ones, should be taken into account in the mass density $\varrho$ of $\mathcal{S}$ in \rfr{qpot}. The resulting constraints on $\kappa$ may, thus, be altered with respect to \rfr{satc}. We will face this issue in a numerical way by integrating the barycentric equations of motion of the Sun, Jupiter, Saturn, Uranus and Neptune modified with the inclusion of the accelerations due to \rfr{Upot}. Moreover, \rfr{qpot} will be calculated by taking into account the contributions of Jupiter, Uranus and Neptune as well. The result is depicted in Figure \ref{plotti}.
}
\begin{figure*}
\centering
\begin{tabular}{c}
\epsfig{file=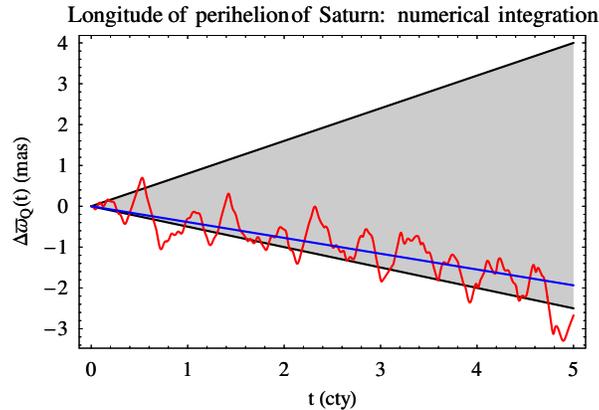,width=0.70\linewidth,clip=}\\
\end{tabular}
\caption{\textcolor{black}{
Gray-shaded area: allowed region  for any  anomalous perihelion precession $\Delta\dot\varpi$ of Saturn according to the constraints in Table \ref{tavolaf}. The black straight lines delimiting it represent the secular perihelion shifts of Saturn corresponding to $\Delta\dot\varpi_{\rm min} = -0.5$ mas cty$^{-1}$ and $\Delta\dot\varpi_{\rm max} = 0.8$ mas cty$^{-1}$ of Table \ref{tavolaf}.
Red curve: time series of the perihelion shift $\Delta\varpi_Q(t)$ of Saturn, in milliarcseconds (mas), due to the MOND planetary quadrupolar potential of \rfr{Upot} caused by the Sun, Jupiter, Uranus and Neptune. It was numerically obtained by simultaneously integrating the equations of motion of the Sun, Jupiter, Saturn, Uranus and Neptune with and without the accelerations induced by $\phi_Q$  over 5 centuries in a Solar System barycentric coordinate system with the ICRF equator as reference $\{x,y\}$ plane. Both the integration shared the same initial conditions which were retrieved from the WEB interface HORIZONS by NASA/JPL. The long time interval of the plot was chosen just for illustrative purposes since it allows to clearly show the secular trend of the perihelion caused by the full MOND planetary quadrupole. The values $\kappa = 3.5\times 10^3,\alpha_{\kappa=1}=1$ were used.
Blue straight line: linear fit of the time series of $\Delta\varpi_Q(t)$. It has a slope as large as $\Delta\dot\varpi_Q = -0.38$ mas cty$^{-1}$, and falls within the gray-shaded allowed region. Larger values of $\kappa$ would yield a MONDian secular trend falling outside it.
}}\lb{plotti}

\end{figure*}
\textcolor{black}{It shows that the inclusion of the other major bodies of the Solar System in the MOND planetary quadrupole of \rfr{qpot} actually enhances its effect on the perihelion of Saturn. Thus, more stringent constraints on $\kappa$ can be inferred:
\eqi |\kappa|\leq 3.5\times 10^3,\lb{newsatc}\eqf which is two orders of magnitude better than \rfr{satc}. Remaining in the Solar System, other authors obtained looser constraints on $\kappa$ from a different class of MOND phenomena occurring in the strong-field regime, i.e. the boundaries of the MOND domains around the zero-gravity points. Bekenstein and Magueijo \cite{2006PhRvD..73j3513B} found $\kappa=1.75\times 10^5$, while Magueijo and Mozaffari \cite{2012PhRvD..85d3527M} inferred $\kappa\gtrsim 1.6\times 10^6$. }

\textcolor{black}{In principle, it may be argued that such  constraints might be optimistic. Indeed, MOND was not included in the dynamical force models which were fitted to the real observations used to produce the INPOP10a ephemerides; thus, the putative MOND signature may have been partly removed from the real residuals in the estimation of, say, the planetary initial conditions. As a consequence, it would be more correct to reprocess the same data record by explicitly modeling the MOND dynamics and determine some dedicated solve-for parameters. On the other hand, it should be considered that, even in such a case, nothing would assure that the resulting constraints on $\kappa$ would necessarily be more trustable than ours. Indeed, it could always be argued that some other mismodelled/unmodeled dynamical feature, either of classical or of exotic nature,  may somehow creep into the estimated MOND parameter(s). 
About the issue of the potential partial removal of an unmodelled signature from the real residuals\footnote{\textcolor{black}{For a recent study explicitly demonstrating such a possibility by fitting certain modified models of gravity to simulated data, see \cite{2012CQGra..29w5027H}. However, its validity could well be limited just to the dynamical models considered and/or to the simulation procedure adopted.}}, it is difficult to believe that it may be a general feature valid in every circumstances for every force models. Otherwise, it would be difficult to realize how Le Verrier \cite{leverrier59} could have positively measured the general relativistic perihelion precession of Mercury \cite{1915SPAW...47..831E} by processing the observations with purely Newtonian models for both the planetary dynamics and for the propagation of light. Here we are not even engaged in measuring some effects; more modestly, we are looking just for upper bounds. As another example, let us consider the Pioneer anomaly \cite{1998PhRvL..81.2858A,2002PhRvD..65h2004A}. In that case, we concluded \cite{2008mgm..conf.2558I} that it could not be due to a gravitational anomalous acceleration directed towards the Sun by comparing the predicted planetary perihelion precessions caused by it with the limits of the anomalous planetary perihelion precessions obtained by some astronomers without explicitly modeling such a putative acceleration. Our conclusions were  substantially confirmed later by dedicated  analyses of independent teams of astronomers. Indeed, either ad-hoc modified dynamical planetary theories were fitted by them to data records of increasing length and quality with quite negative results for values of the anomalous radial acceleration as large as the Pioneer one \cite{2008AIPC..977..254S,2009sf2a.conf..105F,2010IAUS..261..179S,2012sf2a.conf...25F}, or they explicitly modeled and solved for a constant, radial acceleration getting admissible upper bounds \cite{2009IAU...261.0601F} not weaker than those obtained by us \cite{2011JCAP...05..019I}. On the other hand,  Blanchet and Novak \cite{2011MNRAS.412.2530B} inferred their constraints on the EFE-induced MONDian quadrupole effect \cite{2009MNRAS.399..474M} with the same approach followed by us in this paper in obtaining \rfr{satc}: they confronted their analytically calculated perihelion precessions with the admissible ranges for the anomalous precessions obtained by some astronomers without modeling MOND. Finally,} our results support the guess by Milgrom \cite{2012MNRAS.426..673M} that values of $\kappa>10^5$ might be excluded.

The outer planets (Uranus, Neptune, Pluto) are not yet suitable for such kind of analyses: indeed, their orbits are still poorly known because of a lack of extended records of radio-technical data. As far as their perihelia are concerned, their anomalous precessions are constrained to a $4-5$ arcseconds per century ($''$ cty$^{-1}$) level \cite{2010IAUS..261..170P}.
To be more \textcolor{black}{quantitative, a preliminary two-body analysis is adequate for them. F}rom \rfr{rateQ} for moderate eccentricities it turns out
\eqi\kappa\propto  \rp{M_{\rm b}}{a^{7/4}}\ton{\rp{\Delta\dot\varpi}{m}}^{1/2}.\eqf
In addition to Saturn ($m=5.7\times 10^{26}$ kg, $a=9.5$ au), let us consider Pluto ($m=1.3\times 10^{22}$ kg, $a=39.2$ au); Pitjeva \cite{2010IAUS..261..170P} yields $\Delta\varpi=2.84\pm 4.51\ ''$ cty$^{-1}$ for its anomalous perihelion precession.  Thus,
\eqi\rp{\kappa_{\rm Pluto}}{\kappa_{\rm Saturn}}\sim \ton{\rp{m_{\rm Saturn}}{m_{\rm Pluto}}}^{1/2}\ton{\rp{a_{\rm Saturn}}{a_{\rm Pluto}}}^{7/4}\ton{\rp{\Delta\dot\varpi_{\rm Pluto}}{\Delta\dot\varpi_{\rm Saturn}}}^{1/2}\sim 1460;\eqf the constraint on $\kappa$ from Pluto would be 1460 times less tight than \rfr{satc} inferred from Saturn. Although the orbit determination of Pluto will be improved by the ongoing New Horizons mission \cite{2003EM&P...92..477S} to its system, its perihelion precession should be constrained down to a totally unrealistic $0.001$ mas cty$^{-1}$ level in order to yield constraints competitive with \rfr{satc}. An analogous calculation for Neptune ($m=1\times 10^{26}$ kg, $a=30.1$ au, $\Delta\dot\varpi = -4.44\pm 5.40\ ''$ cty$^{-1}$ \cite{2010IAUS..261..170P}) yields $\kappa_{\rm Neptune}\sim 28\kappa_{\rm Saturn}$. It implies that the anomalous perihelion precession of Neptune should be improved down to a $0.1$ mas cty$^{-1}$ level. At present, no missions to the Neptunian system are scheduled. Nonetheless, the OSS (Outer Solar System) mission \cite{2012ExA....34..203C},  aimed to test fundamental and planetary physics with Neptune, Triton and the Kuiper Belt, has been recently proposed; further studies are required to investigate the possibility that, as a potential by-product of OSS, the orbit determination of Neptune can reach the aforementioned demanding level of accuracy.

The situation for Jupiter ($m=1.898\times 10^{27}$ kg, $a=5.2$ au) is, in perspective, more promising. At present, its perihelion precession is modestly constrained at a $-41\pm 42$ mas cty$^{-1}$ level  \cite{2011CeMDA.111..363F}; thus it is currently not competitive with Saturn. A $0.1$ mas cty$^{-1}$ level would be required: such a goal may, perhaps, not be too unrealistic in view of the ongoing Juno mission \cite{2007AcAau..61..932M}, which should reach Jupiter in 2016 for a year-long scientific phase, and of the approved\footnote{\url{http://sci.esa.int/science-e/www/object/index.cfm?fobjectid=50321}} JUICE mission \cite{2011epsc.conf.1343D}, to be launched in 2022, whose expected lifetime in the Jovian system is 3.5 yr.
\subsection{Radial velocities in binaries}\lb{RV}
In general, according to \rfr{dvr}, the most potentially promising binaries are necessarily those orbiting slowly enough to fulfil the quasi-staticity condition. Moreover, they should move in highly elliptical, non-face-on orbits, and their masses should be comparable. Finally, the data records should cover
at least one full orbital revolution.
\subsubsection{Exoplanets}\lb{esopianeta}
The wealth of exoplanets discovered so far allows, at least in principle, to select some of them  for our purposes.

Let us consider 55 Cnc d \cite{2008ApJ...675..790F} which is a Jupiter-like planet ($m\sin I= 3.835 m_{\rm J}$) orbiting a Sun-like star ($M = 0.94 M_{\odot}$; $t_{\rm M}=38$ d) along a moderately elliptic orbit ($e=0.025$) with a period $P_{\rm b}= 14.28\ {\rm yr} = 5218$ d; the other relevant parameters are $\omega=181.3^{\circ}, I=53^{\circ}$. It was discovered spectroscopically; the accuracy in measuring the amplitude $K$ of its radial velocity is \cite{2008ApJ...675..790F} \eqi \sigma_K = 1.8\ {\rm m\ s}^{-1}.\lb{Kv}\eqf
By using \rfr{dvr} for 55 Cnc d and \rfr{Kv}, it turns out
\eqi|\kappa|\leq 7\times 10^8,\eqf which is 3 orders of magnitude weaker than the constraint of \rfr{satc} inferred from the perihelion precession of Saturn.
It should be noticed that the use of \rfr{dvr}, which refers to the shift of the radial velocity over one full orbital revolution, is fully justified since Fischer et al. \cite{2008ApJ...675..790F} analyzed 18 years of Doppler shift measurements of 55 Cnc.

Other wide systems may yield better constraints, although not yet competitive with those from our Solar System. For example, HD 168443c \cite{2011ApJ...743..162P} ($M=0.995M_{\odot}$, $m\sin I= 17.193 m_{\rm J}$, $t_{\rm M}=39.8$ d, $P_{\rm b}=4.79\ {\rm yr}= 1749.83$ d, $e=0.2113$, $a=2.8373$ au, $\omega=64.87^{\circ}$, $\sigma_K=0.68$ m s$^{-1}$) yields
\eqi|\kappa|\leq 3\times 10^7\eqf
by assuming $I=50^{\circ}$. Also in this case the use of \rfr{dvr} is justified since the spectroscopic Doppler measurements cover more than one orbital period. A similar result may occur for 47 Uma d \cite{2010MNRAS.403..731G} ($M=1.03 M_{\odot}$, $m\sin I= 1.6 m_{\rm J}$, $t_{\rm M}=40$ d, $P_{\rm b}=38.3\ {\rm yr}=14002$ d, $e=0.16$, $a=11.6$ au, $\omega=110^{\circ}$, $\sigma_K = 2.9$ m s$^{-1}$), but, in this case, the data used by Gregory et al. \cite{2010MNRAS.403..731G} span a period of just $21.6$ years.
\subsubsection{Spectroscopic stellar binaries}\lb{binaria}
Looking at double lined spectroscopic binary stars, an interesting candidate is the $\alpha$ Cen AB system \cite{2002A&A...386..280P}. It is constituted
by two Sun-like main sequence stars A ($M=1.105M_{\odot}$) and B ($m=0.934M_{\odot}$) revolving  along a wide ($a=23.52$ au) and eccentric
($e=0.5179$) orbit with  $P_{\rm b}=79.91\ {\rm yr}=29187.12\ {\rm d}\gg t_{\rm M}=56.35$ d.  The standard deviations of their radial velocities are  \cite{2002A&A...386..280P} $\sigma_{V_{\rho}^{\rm (A)}}=7.6$ m s$^{-1}$,  $\sigma_{V_{\rho}^{\rm (B)}}= 4.3$ m s$^{-1}$.
Thus, from \rfr{dvr} we obtain the tight constraints
\begin{align}
|\kappa| \lb{kA} & \leq 6.2\times 10^4\ ({\rm A}), \\ \nonumber \\
|\kappa| \lb{kB} & \leq 4.2\times 10^4\ ({\rm B}).
\end{align}
Such bounds are one order of magnitude tighter than the \textcolor{black}{two-body limit}  of \rfr{satc} inferred from the perihelion precession of Saturn\textcolor{black}{, but, on the other hand, the multi-body constraint of \rfr{newsatc} from Saturn's perihelion is better than \rfr{kA}-\rfr{kB} by about one order of magnitude.}

Other aspects of MOND, different from the effect treated here, were investigated with Proxima Centauri\footnote{It should be gravitationally associated with $\alpha$ Cen AB \cite{2006AJ....132.1995W}. Their mutual separation is 15,000 au \cite{2006AJ....132.1995W}.} ($\alpha$ Cen C) \cite{2009MNRAS.399L..21B,2011Ap&SS.333..419B,2012MNRAS.421L..11M}.
\subsection{Pulsars}\lb{pulsar}
In order to fruitfully use \rfr{taup}, the orbital period of the binary chosen should be larger than $t_{\rm M}\approx 46.7$ d, obtained by using the standard value for the pulsar's mass $M=1.4M_{\odot}$; this implies that wide orbits are required. Moreover, they should be rather eccentric as well, and the mass $m$ of the companion should not be too small with respect to the pulsar's one. Finally, timing observations should cover at least one full orbital revolution. As a consequence, most of the currently known binaries hosting at least one radiopulsar are to be excluded because they are often tight systems with very short periods.

A partial exception is represented by the Earth-like planets \cite{2003ApJ...591L.147K} C  ($P_{\rm b}=66.5$ d, $m=0.0163m_{\rm J}$, $a=0.36$ au, $e=0.0186$, $I=53^{\circ}$, $\omega=250.4$) and D ($P_{\rm b}=98.2$ d, $m=0.0164m_{\rm J}$, $a=0.46$ au, $e=0.0252$, $I=47^{\circ}$, $\omega=108.3$) discovered in 1991 around the PSR 1257+12 pulsar ($M=1.4 M_{\odot}$) \cite{2012NewAR..56....2W}; the post-fit residuals for the TOAs was $\sigma_{\delta\tau_{\rm p}}=3.0\ \mu$s \cite{2003ApJ...591L.147K}.
Applying \rfr{taup} to D yields
\eqi|\kappa|\leq 1.5 \times 10^{12}.\eqf
Such a constraints is far not competitive with those inferred from Saturn (Section \ref{pianeta}) and $\alpha$ Cen AB (Section \ref{binaria}).
\section{Summary and conclusions}\lb{sommario}
We looked at the newly predicted quadrupolar MOND effect occurring in non-spherical, isolated and quasi-static $\ton{P_{\rm b}\gg t_{\rm M}=\sqrt{GM_{\rm tot}A_0^{-1} c^{-2}}}$ systems in deep Newtonian regime, and calculated some orbital effects for a localized binary system in the framework of the QUMOND theory.

In particular, we worked out the secular precession of the pericenter, the radial velocity and timing shifts per revolution \textcolor{black}{for a two-body system}. Our results  are exact in the sense that no simplifying assumptions about the orbital geometry were used.


By using the latest orbital determinations of the planets of the Solar System, we inferred $|\kappa|\leq 2.5\times 10^5$ from the supplementary precession of the perihelion of Saturn. \textcolor{black}{Such a bound is based on an expression for the MOND quadrupole which takes into account only the contributions of the Sun and of Saturn itself. Actually, the contributions of the other giant planets of the Solar System do have a non-negligible impact. We evaluated it by numerically integrating the planetary equations of motion. As a result, we found a tighter constraint from Saturn: $|\kappa|\leq 3.5\times 10^3$}. The double lined spectroscopic binary $\alpha$ Cen AB allowed to obtain $|\kappa|\leq 6.2\times 10^4\ ({\rm A}), |\kappa|\leq 4.2\times 10^4\ ({\rm B})$ from our prediction for the shift in the radial velocity. The bounds that can be obtained by extrasolar planets, including also those orbiting pulsars, are not yet competitive. In general, the best candidates are binary systems made of comparable masses moving along accurately determined wide and highly eccentric orbits.

Our constraints are to be intended as somewhat preliminary because, strictly speaking, they did not come from a targeted data processing in which the MOND dynamics was explicitly modeled in processing the real observations and a dedicated solve-for MOND parameter such as $\kappa$ was determined along with the other ones. Nonetheless, they are useful as indicative of the potentiality offered by the systems considered, and may focus the attention just to them for more refined analyses.
%
%
%
%
%
%
%
%
%
%
%

\bibliography{QUMONDbib}{}

\end{document}